# Comparison of threshold-based algorithms for sparse signal recovery

Tamara Koljenšić, Časlav Labudović

Faculty of Electrical Engineering
University of Montenegro
Podgorica, Montenegro
email: koljensict@gmail.com, caslavlabudovic@gmail.com

*Abstract*—**Intensively growing approach in signal processing and acquisition, the Compressive Sensing approach, allows sparse signals to be recovered from small number of randomly acquired signal coefficients. This paper analyses some of the commonly used threshold-based algorithms for sparse signal reconstruction. Signals satisfy the conditions required by the Compressive Sensing theory. The Orthogonal Matching Pursuit, Iterative Hard Thresholding and Single Iteration Reconstruction algorithms are observed. Comparison in terms of reconstruction error and execution time is performed within the experimental part of the paper.**

## I. Introduction

The novel theory of Compressive Sensing (CS) is an intensively growing approach in signal processing. Traditional approach for signal reconstruction using acquired samples, follows the Shannon-Nyquist sampling theorem. For signals with high maximal frequency Shannon-Nyquist sampling procedure can result in a large number of samples. Against the traditional perspective in data acquisition, the CS approach allows exact signal reconstruction using small set of randomly acquired samples. The CS offers the possibility to acquire less data then it is commonly done, but still to be able to reconstruct the entire information. This approach attracted much interest in the research community and found wide-ranging applications from astronomy, biology, communications, image and video processing, medicine, to radar. Compressive sensing opens the possibility for simplifying the acquisition devices and apparatus, reducing the number of sensors, acquisition time, and storage capacities.

To enable efficient recovery by using the CS approach, the signals have to satisfy certain conditions such as sparsity and incoherence. Sparsity is one of the main requirements that should be satisfied. It implies that the signal in different domains: time, frequency or time-frequency domains has only small number of non-zero coefficients. If the incoherence property is satisfied, the reconstruction using small set of samples is assured.

Signal recovery is based on a powerful mathematical algorithms for error minimization. Some of them, like basis pursuit, Dantzig selector and gradient-based algorithms rely on linear programming methods. As much as they're accurate, they are computationally demanding and not always suitable in practical applications. Many alternative approaches, like greedy algorithms – Matching Pursuit and Orthogonal Matching Pursuit are being developed to reduce the computational complexity. Also, recently proposed threshold based algorithms provide high reconstruction accuracy with low computational complexity: (e.g. Iterative Hard Thresholding - IHT, Iterative Soft Thresholding - IST).

In this paper the focus is on the commonly used threshold-based algorithms. The comparison of the several algorithms is done. The sinusoidal multicomponent signal is observed, which can be defined by using the following relation:

$$x(n) = \sum_{k=1}^{K} a_k e^{-j2\pi kn}, \ n \in [0, N-1], \tag{1}$$

where $K$ denotes number of sinusoidal components in observed signal, while $N$ is signal's length. Usually, these signals have large number of samples $N$ but small number of non-zero components in frequency domain ($K \ll N$). The reconstruction of such signals using various solutions has been presented in this work. Noiseless signal cases are examined.

The organization of this paper is as follows. In Section II, the fundamental concepts of CS theory and signal reconstruction method are given. The overview of the applied CS algorithms is given in the Section III. The CS application to band-limited sparse signals is discussed in Section IV. Concluding remarks are given in Section V.

## II. Basics of the Compressive Sensing approach

In order to reconstruct the signal with high accuracy, traditional methods require sampling frequency to be twice the maximal signal frequency. This leads to a high number of signal samples that are acquired and stored. Beside the large number of samples, another problem is presence of noise, which can lead to missing signal information. New theory, the CS theory, overcomes those limits. It directly senses the data in a compressed form – i.e. at a lower sampling rate, and allows recovering the intentionally omitted or missing signal samples.

Let $\mathbf{f}$ be an $N$-dimensional signal of interest, which is sparse in the transform domain represented with the transformation matrix $\mathbf{\Psi} \in R^{N \times N}$. The sparse representation of $\mathbf{f}$ over the basis $\mathbf{\Psi}$ is represented by the vector $\mathbf{p}$. Then $\mathbf{f}$ can be given by

$$\mathbf{f} = \mathbf{\Psi}\mathbf{p} \ . \tag{2}$$

If $\mathbf{\Psi}$ is e.g. the inverse Fourier transform (FT) matrix, then $\mathbf{p}$ can be regarded as the frequency domain representation of the time domain signal, $\mathbf{f}$. Signal $\mathbf{f}$ is said to be $K$-sparse in the $\mathbf{\Psi}$ domain if there are only $K (K << N)$ out of $N$ coefficients in $\mathbf{p}$ that are non-zero. If a signal is able to be sparsely represented in a certain domain, the CS technique can be invoked to take only a few linear and non-adaptive measurements.

The set of random measurements are selected from signal $\mathbf{f}(N \times 1)$, which can be written by using random measurement matrix $\mathbf{\Phi}(M \times N)$ as follows:

$$\mathbf{d} = \mathbf{\Phi}\mathbf{f} \tag{3}$$

From (2) and (3):

$$\mathbf{d} = \mathbf{\Phi}\mathbf{\Psi}\mathbf{p} \tag{4}$$

The incoherence is another important condition that basis matrix $\mathbf{\Psi}$ and measurement matrix $\mathbf{\Phi}$ should satisfy to make the CS reconstruction possible. The relation between the number of nonzero samples in the transform domain $\mathbf{\Psi}$ and the number of measurements (required for reconstruction) depends on the coherence between the matrices $\mathbf{\Psi}$ and $\mathbf{\Phi}$. More specifically, a good measurement will pick up a little bit of information of each component in $\mathbf{p}$ based on the condition that $\mathbf{\Phi}$ is incoherent with $\mathbf{\Psi}$. As a result, the extracted information can be maximized by using the minimal number of measurement.

### III. ALGORITHMS FOR SPARSE RECOVERY

In this section, some of the commonly used threshold-based algorithms are described. The subjects of this analysis are Orthogonal Matching Pursuit (OMP), Iterative Hard Thresholding algorithm (IHT) and Single Iteration Reconstruction Algorithm (SIRA).

Knowing CS sensing matrix $\mathbf{\Omega} = \mathbf{\Psi}\mathbf{\Phi}$ and measurement vector $\mathbf{d}$, the OMP algorithm approximates signal $\mathbf{f}$ as linear combination of columns in $\mathbf{\Omega}$. In each iteration, a set of columns is expanded with additional column that correlates best with the residual signal. The algorithm terminates until residual falls below determined threshold. OMP can be summarized as follows:

1. Variables initialization: set the approximation error $\mathbf{r}^0 = \mathbf{d}$, the initial solution to $\mathbf{f}^0$ and $\mathbf{S}^0 = \emptyset$.

2. Do following steps till the stopping criterion is met:

  a) $\quad \mathbf{S}_n = \mathbf{S}_{n-1} \cup \arg_i \max \left| \langle \mathbf{r}_{n-1}, \mathbf{\Omega}_i \rangle \right| \ , \tag{5}$

  b) $\quad \mathbf{f}_n = \arg \min_f \left\| \mathbf{r}_{n-1} - \mathbf{S}_{n-1}\mathbf{f}_{n-1} \right\|_2^2 \ , \tag{6}$

  c) $\quad \mathbf{r}_n = \mathbf{r}_{n-1} - \mathbf{S}_n \mathbf{f}_{n-1} \ , \tag{7}$

  d) $\quad n = n+1$ and $\mathbf{S}_n = \mathbf{S}_{n-1} \cup \arg_i \max |\langle \mathbf{r}_{n-1} - \mathbf{\Omega}_i \rangle| \ (8)$

  until $n \le K$, where $K$ is number of signal components.

The IHT algorithm [12] is an iterative algorithm which uses non-linear operator to reduce the $l_0$ - norm in each iteration. The algorithm solves the following problem:

$$\min_f \left\| \mathbf{d} - \mathbf{\Omega}\mathbf{f} \right\| + \lambda \left\| \mathbf{f} \right\|_0 \tag{9}$$

From the optimization problem described by (9), the following iterative algorithm is derived. The non-linear operator is denoted as $H_k(a)$ and sets all but the largest (in terms of magnitude) $k$ elements of $a$ to zero. For given $\mathbf{f}^0$, the algorithm iterate:

$$\mathbf{f}^{n+1} = H_k(\mathbf{f}^n + \mathbf{\Omega}^T(\mathbf{d} - \mathbf{\Omega}\mathbf{f}^n)), \tag{10}$$

Until either $k > N_{max}$ or $\left| \mathbf{d} - \mathbf{\Omega}\mathbf{f}^n \right|_2 < \varepsilon$   $H_k$ is defined as follows:

$$H_k(f_i) = \begin{cases} 0, |f_i| < k \\ f_i, |f_i| > k \end{cases} \tag{11}$$

The Single Iteration Reconstruction Algorithm - SIRA is based on threshold calculation and choice of initial discrete Fourier transform (DFT) components that are above defined threshold. Initial DFT is calculated based on a set of available signal samples. It is shown that components above the threshold correspond to the signal components, while the components below the threshold are considered as noise. We make our analysis on the assumption that some random samples are omitted. The noise appears in signal and variance of it can be modeled as:

$$\text{var} = M \frac{N_a}{N-1}(A_1^2 + A_2^2 + ... + A_k^2), \tag{12}$$

where M is the number of missing samples, while $N_a$ is the number of available samples. This variance will be used in threshold calculation:

$$T = \frac{1}{N}(-\text{var}^2 \log_{10}(1 - \sqrt[N]{P})^{1/2} \tag{13}$$

The P is the probability that (N-K) components that correspond to noise, are lower than the threshold. The samples positions in the initial DFT that are above the threshold are used for the calculation of the exact DFT signal amplitudes, while the other frequency positions are filled with zeros. Vector of the initial DFT is formed using the available time domain signal samples, i.e. using the vector of measurements. Let v($M \times 1$) denotes the measurement vector. The initial DFT vector V is therefore formed as:

$$V(f) = \sum_{a=1}^{N_a} v(a) e^{-j2\pi f a/N}, f = 1,...,N \qquad (14)$$

Positions of the components above the threshold are obtained using following relation:

$$pos = \arg\{|V| > T\}. \qquad (15)$$

We found frequency positions, but to find the exact amplitudes on those positions requires solving minimization problem:

$$\mathbf{X} = (\mathbf{A}^*_{CS}\mathbf{A}_{CS})^{-1}(\mathbf{A}_{CS} v) \qquad (16)$$

where CS matrix $\mathbf{A}_{cs}$ is formed as $\mathbf{A}(P_v, pos)$, $\mathbf{A}^*_{cs}$ is Hermitian transpose of the matrix $\mathbf{A}_{cs}$ and $\mathbf{P_v}$ denotes vector of the available signal samples positions.

## IV. CS RECONSTRUCTION APPLIED ON SPARSE BAND-LIMITED SIGNAL

In this section, the three described algorithms, OMP, IHT and SIRA, are tested on the sparse band-limited signal, consisted of 7 components. The components' magnitudes differ significantly from component to component. The signal is described using the following relation:

$$x = A_1 e^{j2\pi 32n/N} + A_2 e^{j2\pi 38n/N} + A_3 e^{j2\pi 130n/N} + A_4 e^{j2\pi 148n/N} \dots$$
$$+ A_5 e^{j2\pi 272n/N} + A_6 e^{j2\pi 415n/N} + A_7 e^{j2\pi 435n/N}$$
$$(17)$$

where component magnitudes are:

$A_1 = 3.5; A_2 = 3; A_3 = 1.75; A_4 = 2.5; A_5 = 3.75; A_6 = 2.3; A_7 = 3.3;$

an N is the length of the signal and n ∈ (1, $N$). Signal is considered as sparse in the DFT domain.

In terms of the reconstruction error and the execution time, the same number of measurements is used: 200 samples (39.06% of the total signal length), 225 (43,94% of the total signal length), 250 (48,82% of the total signal length), 275 (53,71 % of the total signal length) and with M=300 samples (58,59% of the total signal length). Fig. 1a shows time and Fig. 1b shows DFT domain of the original signal.

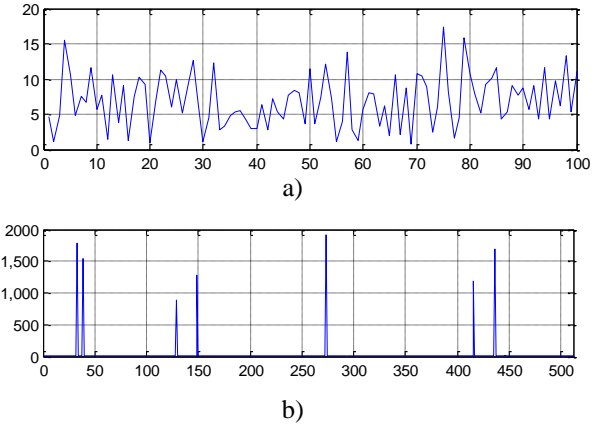

Figure 1: a) Zoomed time domain of the original, b) Fourier domain of the original signal

Table 1 shows the error and reconstruction time for the obtained measurements. The error is defined as a maximum magnitude difference between the original DFT and the reconstructed one.

Table 1: Error and reconstruction time for different numbers of samples

| Number of samples M | ALGORITHM | SIRA | OMP | IHT |
|---|---|---|---|---|
| *200* | **TIME(sec)** | 0.013945 | 0.098406 | 0.055230 |
| | **ERROR** | $3.5391 \cdot 10^{-11}$ | 155.1299 | $6.698 \cdot 10^{-6}$ |
| *225* | **TIME(sec)** | 0.013577 | 0.076467 | 0.067002 |
| | **ERROR** | $3.5391 \cdot 10^{-11}$ | 81.3499 | $5.8892 \cdot 10^{-6}$ |
| *250* | **TIME(sec)** | 0.016528 | 0.095628 | 0.068634 |
| | **ERROR** | $3.5391 \cdot 10^{-11}$ | 70.1819 | $7.073 \cdot 10^{-6}$ |
| *275* | **TIME(sec)** | 0.014068 | 0.089988 | 0.063323 |
| | **ERROR** | $3.5391 \cdot 10^{-11}$ | 9.7435 | $6.263 \cdot 10^{-6}$ |
| *300* | **TIME(sec)** | 0.018851 | 0.103983 | 0.059231 |
| | **ERROR** | $3.5391 \cdot 10^{-11}$ | 8.6813 | $6.133 \cdot 10^{-6}$ |

From the Table 1 it can be seen that the SIRA algorithm is the fastest with the most accurate reconstruction of the original signal. For M between 200 and 300, the minimum error was obtained using SIRA.

Fig.2 shows the different reconstructions for the minimum number of measurements

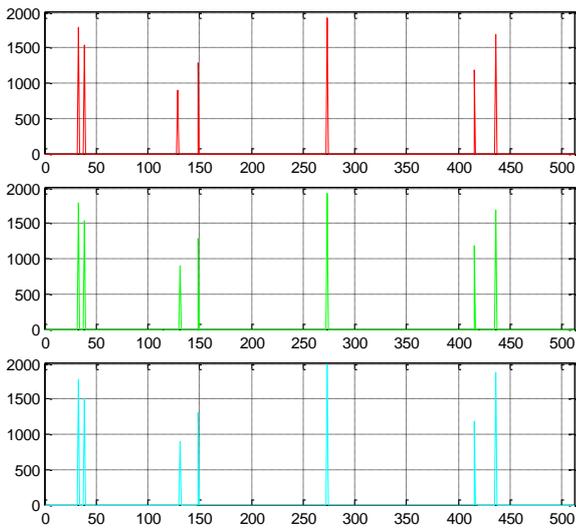

Figure 2: Comparison of the different algorithms reconstruction for: SIRA (M=170), IHT (M=34) and OMP (M=200), from the top to the bottom

OMP algorithm was able to guess each position of samples more frequently than SIRA but with less accuracy. IHT could reconstruct the original signal perfectly, but with the information about expected number of components.

Fig.3 shows IHT reconstruction when the number of components is larger than predicted.

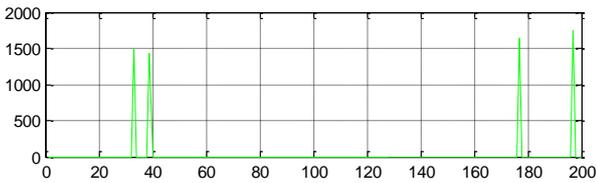

Figure 3: IHT reconstruction in case of unknown expected components

Fig.4 shows error for the minimal measurement numbers needed to successfully reconstruct original signal

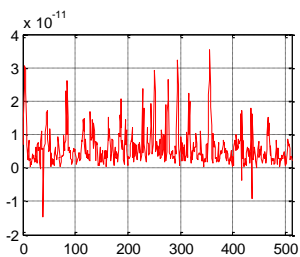

a) SIRA for M=170;

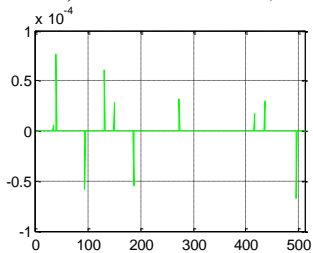

b) IHT for M=34

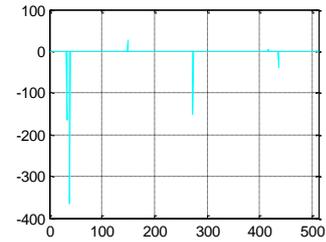

c) OMP for M=200

Fig. 5a shows the error for all three algorithms for number of measurements range from 200 to 300 and Fig. 5b shows zoomed error for SIRA and IHT

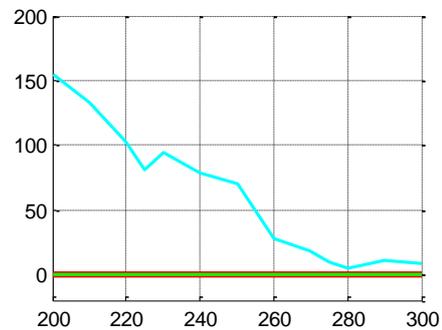

a)

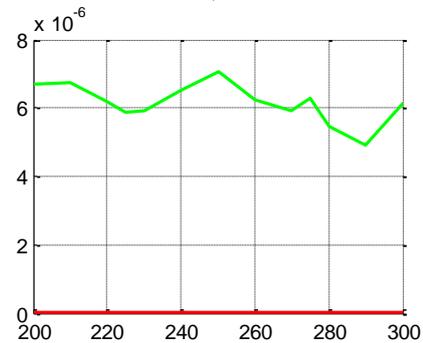

b)

Figure 5: a) Error for the measurements between 200 and 300
b) Zoomed errors for SIRA, IHT

## V. CONCLUSION

Performance of the several reconstruction algorithms applied on sparse band-limited signal are considered in the paper. Signal is being reconstructed changing the number of measurements.

Comparing maximum error and execution time, the best results were obtained by using SIRA. It is important to acknowledge that OMP was able to find all samples' positions in the same number of iterations as SIRA but error was significantly bigger. IHT algorithm demands a priori knowledge of the signal, as one of his inputs is the number of components we are searching for. In case of familiar signal, IHT used 6, 65% of the signal length to reconstruct signal successfully. That's 20% of SIRA's requirements but SIRA can work without previous signal's background.